# Fabrication of highly uniform laser-induced periodic structures on polycarbonate via UV femtosecond pulses

Matina Vlahou[1,2], Nektaria Protopapa[1,3], Stella Maragkaki[1], George D. Tsibidis[1] and Emmanuel Stratakis[1,3, *]

[1] Institute of Electronic Structure and Laser (IESL), Foundation for Research and Technology (FORTH), N. Plastira 100, Vassilika Vouton, 70013, Heraklion, Crete, Greece
[2] Department of Materials Science and Technology, University of Crete, 71003 Heraklion, Crete, Greece
[3] Department of Physics, University of Crete, 71003 Heraklion, Crete, Greece

*stratak@iesl.forth.gr

# Abstract

In this work, we focus on the fabrication of highly uniform laser-induced periodic surface structures (LIPSS) on bulk polycarbonate (PC) using 258 nm femtosecond laser pulses. A systematic approach was pursued to investigate the influence of various laser parameters such as fluence, effective number of pulses, energy dose and polarisation on the features of the generated LIPSS. Experimental results showed that linearly polarized beams produce LIPSS with period comparable to the laser wavelength. Moreover, it was observed that the orientation of LIPSS is either parallel or perpendicular to the laser polarization, depending on the excitation level. These features are similar to the LIPSS formed on dielectrics, despite the considerably higher absorption and larger extinction coefficient (~$10^{-2}$) of polycarbonate at 258 nm. The orientation and features of the patterns at different excitation levels are explained using Sipe's theory, while the excitation and induced carrier densities were quantified through the application of a theoretical physical model that describes ultrafast dynamics in polymers. Furthermore, observations revealed that low excitation levels with linearly polarized enhance the uniformity of LIPSS, whereas irradiating PC targets with circularly polarized beams leads to the formation of microscale topographies decorated with protruding nanosphered-like structures, with their size varying based on the laser conditions. A detailed analysis of the impact of the induced topographies on their wetting and optical properties demonstrated a steady increase in hydrophilicity over time, in contrast to a modest increase in optical absorbance. The results regarding the morphological features and properties of the induced topographies on PC are anticipated to support future efforts in creating patterned surfaces on polymeric materials for potential applications.

# 1. Introduction

Surface functionalization of polymers has recently received considerable attention due to the important applications including tissue engineering and drug delivery as a consequence of their high biocompatibility [1]. Among the various approaches employed towards generating application-based topographies such as nanoimprint lithography [2] and micro-contact printing [3], laser processing with ultrashort pulses constitutes a precise, single-step and scalable method to fabricate highly ordered, and complex multiple length scale patterns down to nanoscale, without using toxic chemicals [4]. In particular, the capability to fabricate complex topographies on polymeric materials is expected to benefit the performance of next-generation flexible electronic devices for biomedical and photonic applications. As a result, the employment of ultrashort pulsed lasers to fabricate functional polymeric surfaces appears to be considerably superior to those obtained via alternative methods.

Among the wealth of the surface structures generated by ultrashort laser pulses, the so-called laser-induced periodic surface structures (LIPSS) on various solids have been explored extensively to understand their origin (see Refs. [[4, 5]] and references therein). In particular, low-spatial frequency LIPSS (LSFL) are assumed to be generated from the interference of the incident laser beam with laser-induced surface scattered waves. In regard to the manifestation of LIPSS formation with femtosecond (fs) pulses, two different types of LSFL structures are reported in the literature for weak absorbing materials such as polymers or dielectrics: LSFL structures with orientation parallel to that of the electric field of the incident beam (termed here as LSFL-II) have been predicted and observed for moderate intensities. The size of LSFL-II is about $\lambda_L/n$ (or larger), where $\lambda_L$ and $n$ stand for the laser wavelength and the refractive index of the material, respectively [6-9]. By contrast, LSFL structures with orientation perpendicular to that of the electric field of the incident beam (termed here as LSFL-I) have also been observed in strong absorbing materials (i.e. metals and semiconductors) [10-12]. The commonly accepted mechanism that accounts for the formation of LSFL-I is the interference of the incident beam with excited Surface Plasmon Polaritons (SPP) [5, 13, 14]. Nevertheless, even for weak absorbing materials, intense laser beams can eventually lead to excitation levels at which a transition of the dielectric material to a 'metallic' state can be achieved generating SPPs. Thus, if sufficient excitation levels are reached, appropriate conditions are fulfilled for the formation of LSFL-I of sizes about $\sim\lambda_L$ [6, 7, 9, 15]. In principle, the irradiation conditions, surface wave characteristics and dielectric permittivity of the material determine predominantly the type of induced topography.

On the other hand, in previous reports, it was shown that the laser polarization plays a dominant role in the fabrication of 1D- or 2D- LIPSS on various materials (see Ref. [4] and references therein). It is noted that the widely used terms '1D'- or '2D'-LIPSS [16] do not indicate that the shape of the formed structures has one (for '1D-LIPSS') or two (for '2D-LIPSS') dimensions. All the induced objects are three-dimensional structures and 1D-LIPSS consist of parallel ripples which are formed by irradiating the surface with linearly polarized laser pulses (i.e. LSFL-I or LSFL-II) while 2D-LIPSS are surface patterns characterized by periodicity along two or more directions (e.g. square or triangular or dot-like structures, produced through the employment of circularly polarized laser pulses or crossed polarized double pulses with time delays in the picoseconds range [17, 18]).

Despite the extensive studies on LIPSS on various solids, there exist only a few reports on LIPSS formation on polymeric materials upon exposure of the solids to fs laser pulses [1, 9, 19, 20]. Most works have focused on the fabrication of periodic patterns with either nanosecond or picosecond pulsed laser sources [21, 22], while a limited exploration has been conducted of the distinct challenges associated to ultraviolet (UV) fs laser-based processing. In a recent report [23], experimental results on nanostructuring of one type of polymeric material (i.e. polysterene thin films) with fs UV pulses were presented, however, our knowledge of the underlying physical mechanisms for the LIPSS formation, the correlation of the features of the produced patterns (i.e. LIPSS orientation, sizes, shapes) with the laser parameters and the influence of the polarization state (i.e. linear or circular) on the induced topographies following irradiation of polymeric materials with fs UV pulse was still insufficient.

On the other hand, patterning with UV fs laser sources appears to be a challenging aspect due to the pronounced enhanced absorptivity of the polymers at this spectral region [24]; more specifically, polycarbonate (PC) which is the focus of this work exhibits a UV- absorptivity that increases due to photo-degradation [25]. Furthermore, a range of UV wavelengths where patterning of PC has yet to be explored is those below 275 nm which are characterised by significantly higher extinction coefficient $k$ (~$10^{-2}$) compared to the value at longer wavelengths (>$10^{-4}$) [24, 26]. Thus, the investigation of patterning of PC at a laser wavelength that yields relatively enhanced absorptivity to the material is significant as this could reveal a series of fascinating physical phenomena related to the excitation process. Additionally, when comparing the behavior of polycarbonate and dielectrics at short laser wavelengths, the high absorption of PC is expected to result in a substantial amount of energy being absorbed, which is critical for laser processing. In contrast, dielectrics exhibit low absorption and do not undergo significant changes unless exposed to extremely high light intensities. Thus, although similarities in LIPSS features on polymeric and dielectric materials have been also observed at higher wavelengths [1, 6, 8, 9, 19, 20, 27, 28], a crucial question arises that is related to whether LIPSS on PC with fs pulses exhibit characteristics similar to those formed in materials with higher absorptivity, such as semiconductors or highly conductive materials (i.e. metals).

To address the aforementioned research gaps, we have performed a detailed investigation of the LIPSS formation on bulk PC using 258 nm fs laser pulses. To evaluate the influence of the laser polarisation on the topographies attained, results for linearly and circularly polarised beams are presented and compared with observations reported in previous reports in dielectric materials at higher photon energies. We, also, explore the impact of other laser parameters such as the fluence, effective number of pulses and energy dose on the size and orientation of LIPSS. The excitation levels and electron dynamics at various fluences are quantified using a theoretical model for polymers [29] as this approach aims to offer a systematic methodology for pattern fabrication. The role of the induced patterned surfaces on the wetting and optical properties was also analysed and discussed which is important for various applications.

## 2. Results and Discussion

A parametric study has been conducted to investigate and analyze the LIPSS structures formation on PC surfaces in various laser conditions. The aim is to correlate the different types of LIPSS formed with the incident laser fluence, energy dose, and laser polarization state. The parametric analysis is complemented with a study on the influence of the topographical features on the wetting and optical properties of the patterned surfaces. A detailed description of the experimental set up is presented in the Materials and Methods Section.

## 2.1 Surface processing of PC with UV fs pulses

### *2.1.1 Generation of topographies using linearly polarized UV fs pulses*

A detailed analysis of the LIPSS topographies produced with linearly polarised beams at various (peak) fluences $F$ and effective number of pulses, $N_{eff}$, was performed. It is observed that both the fluence and the energy dose influence the orientation, size and uniformity of the induced periodic structures (see Table 1). The parametric study reveals morphological changes associated with excitation levels and thermal effects, depending on the various irradiation parameters. In regard to the influence of the underlying physical phenomena in the features of the induced structures, previous reports have shown that electrodynamical effects and electron dynamics predominantly dictate the formation, orientation and periodicity of LIPSS; by contrast, thermal effects mainly affect the distribution of height of the produced patterns (see [4] and references therein). The investigation of the thermal phenomena is not within the scope of the current study.

It is recalled that in contrast to other solids, the production of LIPSS on polymers requires irradiation with several tens to some thousands of pulses of very low fluence to generate well-defined patterns [9], [19]. In particular, using a low incident fluence of $F=8$ mJ/cm$^2$, it is observed that for a number of effective pulses equal to $N_{eff}=3125$, LSFL-II structures start to form that become more pronounced upon increasing $N_{eff}$ (Fig.1). It is evident that by increasing the excitation level (at higher fluences of $F=10$ mJ/cm$^2$ and 13 mJ/cm$^2$) the formation of LIPSS can be accomplished at even smaller $N_{eff}$ (=250). Finally, at the highest fluence $F=13$ mJ/cm$^2$ and $N_{eff}=3125$ used, besides the apparent disintegration of the LIPSS, the formation of nanosphered structures is observed.

| LIPSS type | Orientation of LIPSS to polarization | Fluence (mJ/cm$^2$) | Number of pulses | Uniform LIPSS |
|---|---|---|---|---|
| LSFL-II | parallel | 8 | 3125-5000 | √ |
| LSFL-II | parallel | 8 | 8064-12500 | × |
| LSFL-II | parallel | 10-13 | 250-1000 | √ |
| LSFL-II | parallel | 10-13 | 2000-3125 | × |
| LSFL-II | parallel | 30-80 | 5-30 | × |
| LSFL-II | parallel | 130-250 | 200 | √ |
| LSFL-II | parallel | 100-410 | 5-20 | √ |
| LSFL-II | parallel | 250 | 100 | × |
| LSFL-I | perpendicular | 600 | 15-26 | × |

**Table 1: LIPSS processing parameters.** Summary of laser parameters and corresponding types of LIPSS formed on PC using linearly polarised beams.

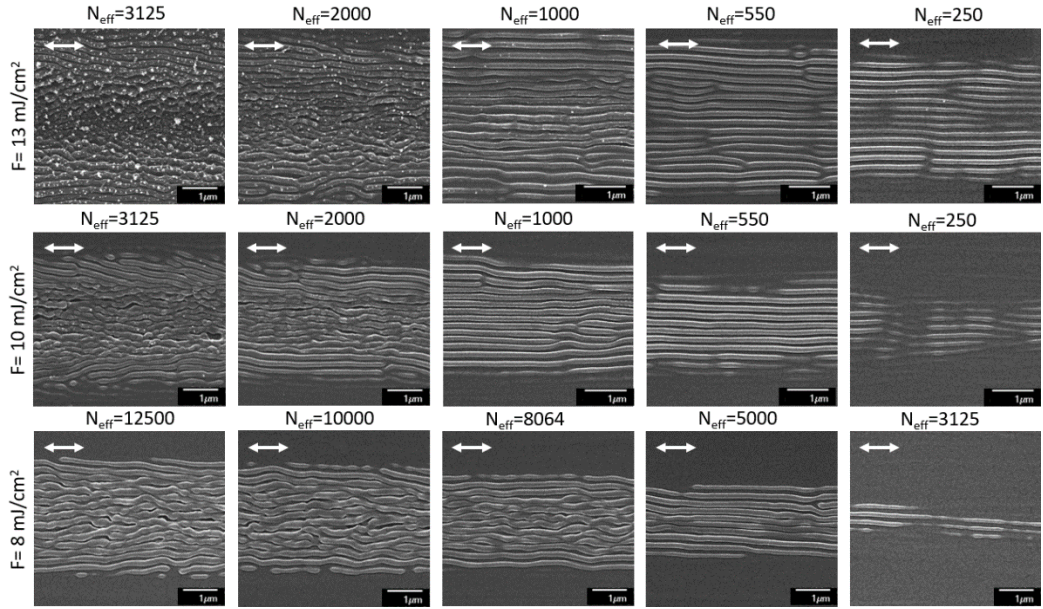

**Figure 1. LSFL-II evolution.** SEM images of PC surfaces illustrating the LSFL-II structures evolution upon irradiation at various fluences and effective number of pulses. The *white* double-headed arrow indicates the direction of the laser polarisation.

Another interesting observation is that upon increasing $N_{eff}$, the uniformity of the LIPSS deteriorates (see for example in Fig. 1 the decrease in uniformity at (i) $F=8$ mJ/cm$^2$, $N_{eff}=8064$, (ii), $F=10$ mJ/cm$^2$, $N_{eff}=2000$, (iii) $F=13$ mJ/cm$^2$, $N_{eff}=1000$). To relate the gradual disappearance of the uniformity with the subsequent observation of nanosphered structures at even higher $N_{eff}$ and $F$ values, one could argue that as the energy dose increases, the pronounced thermal effects lead to the formation of voids; the presence of such defects, whose size and shape is expected to vary upon increasing

$N_{eff}$ [28], potentially influences both the laser energy absorption and distribution which eventually affect the LIPSS uniformity. The impact of the combination of $F$ and $N_{eff}$ on the uniformity of LSFL-II was also explored at higher energy values. As shown in Fig.2a, the uniformity can be preserved at higher fluences provided that a lower $N_{eff}$ value is used, indicating the crucial role of the energy dose used.

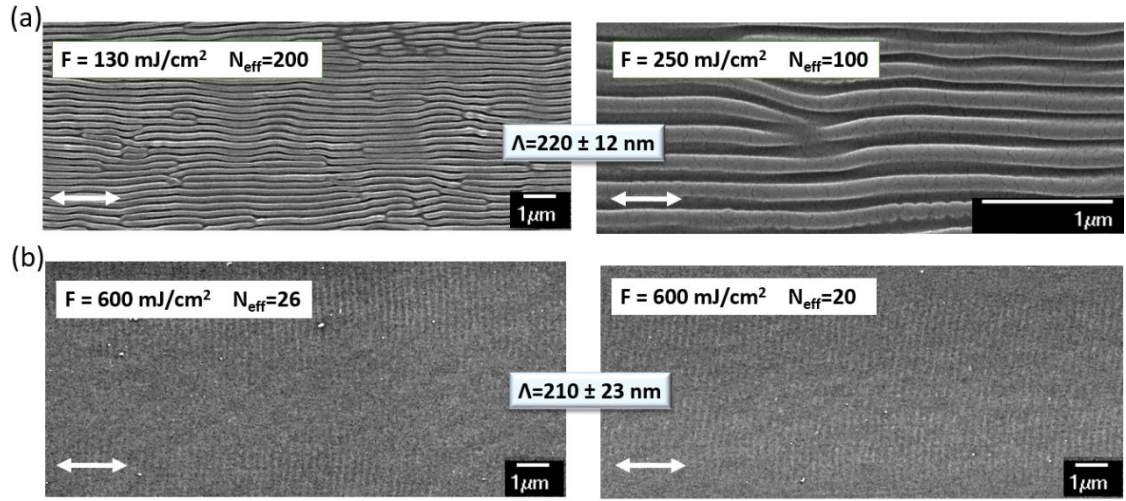

**Figure 2**. **LSFL-I and LSFL-II structures.** Scanning Electron Microscopy (SEM) images of LIPSS on PC formed upon irradiation with linearly polarized UV fs laser pulses row (a) LSFL-II structures and row (b) LSFL-I structures. The *white* double-headed arrow indicates the direction of the laser polarisation.

Remarkably, irradiation at even higher fluences leads to the formation of LSFL-I structures which are oriented perpendicular to the laser polarisation (Fig.2b), although less uniform than the LSFL-II ones.

All the above results are summarized in the morphological maps illustrated in Figs.3a, b, respectively. The first map (Fig.3a) reveals that LSFL-II structures start to appear at a fluence close to the damage threshold (see Supplementary Material) and for a low $N_{eff}$ (up to 10 pulses). These periodic structures are termed as '*Near damage threshold LSFL-II*' (*green* rhombus in Fig.3a). At higher fluence values and up to 600 mJ/cm$^2$ and at $N_{eff}$ =5-30, '*Non-uniform LSFL-II*' structures are formed (*purple* square in Fig.3a). On the other hand, at fluences close to 600 mJ/cm$^2$ and at $N_{eff}$ =15-26, '*LSFL-I*' are produced (*yellow* triangles in Fig.3a). Finally, at fluences greater than 200 mJ/cm$^2$ and at $N_{eff}$>30 no periodic pattern is formed; nevertheless, to characterize the induced surface modification in that regime, the produced structures have been classified as '*Ablation-Roughness*' (*blue* rhombi in Fig.3a).

The second morphological map (Fig.3b) focuses mostly on the influence of *higher $N_{eff}$* on the topographical features observed. Results show that LSFL-II start to appear at, approximately, $F$=8 mJ/cm$^2$ and $N_{eff}$ = 3000. These are the so-called '*Near damage threshold LSFL-II*' (*green* rhombi in Fig.3b). At this fluence and above ~8000 pulses, the formation of '*Non-uniform LSFL-II*' structures is manifested (*purple* triangles in

Fig.3b). Similarly, as shown in Fig.1 and Fig.3c, at a relatively comparable fluence, the formation of '*Non-uniform LSFL-II*' structures is possible at even lower $N_{eff}$. By contrast, there is a small window for the production of '*Uniform LSFL-II*' (*orange* squares in Fig.3b) where fluences up to 250 mJ/cm$^2$ and $N_{eff}$ ranging from ~100 to ~1000 lead to homogeneous LSFL-II. Experimental observations, also, manifest the formation of such type of structures at even higher $N_{eff}$ ($N_{eff}$=5000). Finally, at large excitation levels ($N_{eff} \geq 4000$, $F \geq 0.4$ J/cm$^2$), micron-sized '*Grooves*' (*black* rhombi in Fig.3b) are generated. Such supra-wavelength structures have also been observed at high excitation levels in weak [30] and strong absorbing materials [31-33] and they are oriented parallel to the laser beam polarization.

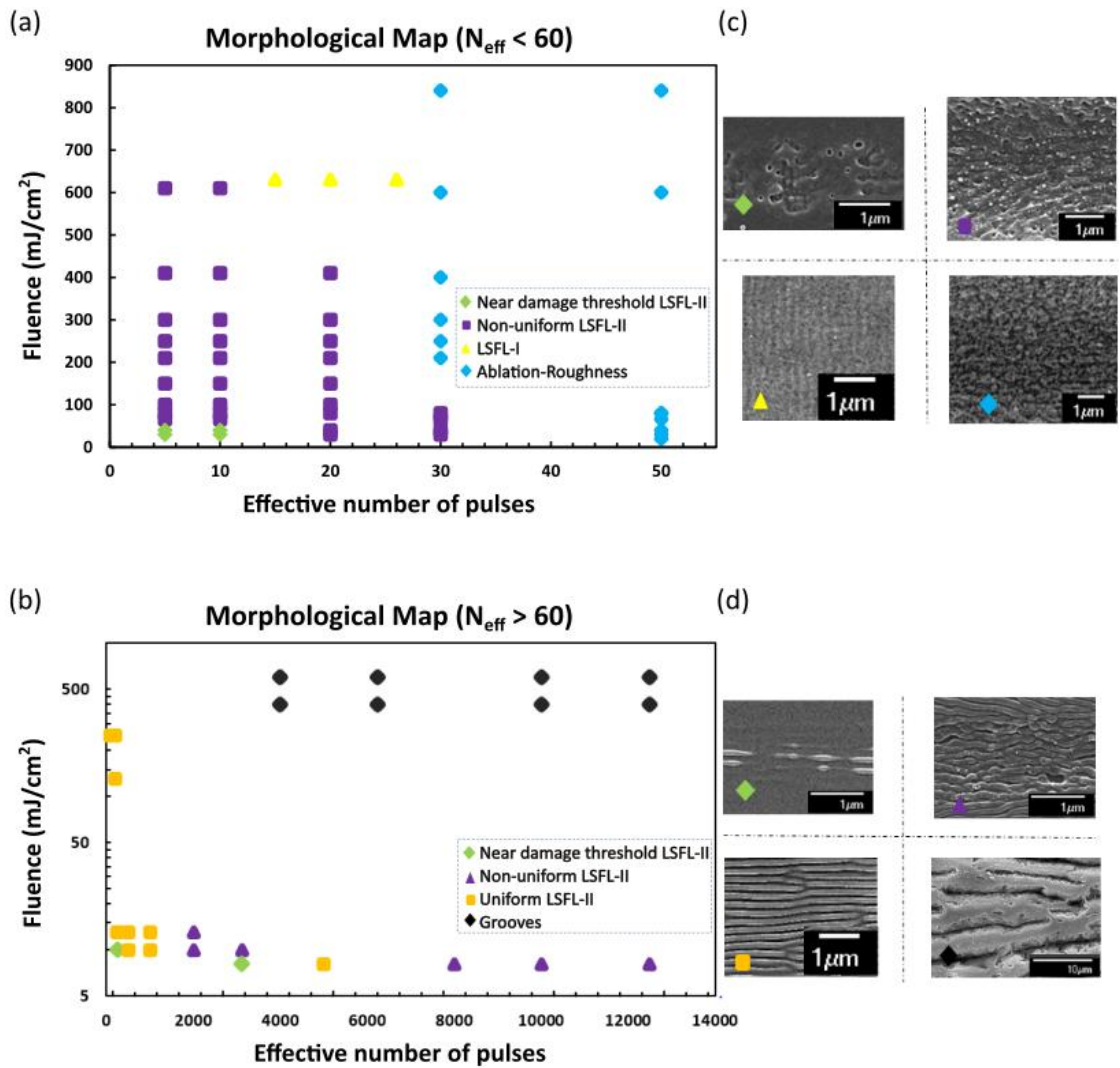

**Figure 3**. **Morphological map of the characteristic surface morphologies.** Maps obtained for (a) $N_{eff}$<60, (b) $N_{eff}$>60. In (c) and (d), representative SEM images of the different topographies obtained in various conditions are illustrated. In all cases, the laser polarization direction is horizontal.

Apart from the identification of the various topographies attained (Fig.3), we performed a systematic analysis aimed to derive the dependence of the LIPSS periodicities on $N_{eff}$ and $F$. A first estimation of the LIPSS frequencies shown in Fig.2 is that LSFL-I and LSFL-II structures exhibit comparable periodicities, namely $\Lambda$=210 nm and $\Lambda$=220 nm, respectively.

The LIPSS period values were obtained following a Fast Fourier Transform (FFT) analysis of the respective SEM images. Figs.4a,b show the dependence of the generated LIPSS periods as a function of $N_{eff}$ at 10 mJ/cm$^2$ and 8 mJ/cm$^2$, respectively. Despite the relatively large error in the determination of $\Lambda$ ($F$=10 mJ/cm$^2$), results demonstrate a LIPSS period that appears to remain, generally, unchanged (i.e. independent on the number of pulses). At lower fluences ($F$=8 mJ/cm$^2$), a practically constant periodicity at increasing $N_{eff}$ is also observed (Fig.4b). Certainly, more investigation and further reduction of the uncertainty in the evaluation of LIPSS periodicities are required including the evaluation of the period at lower energy doses to deduce whether the observed unchanged LIPSS periodicity occurs at any value of $N_{eff}$; the latter discussion is generated from the behaviour manifested by other wide band gap materials (i.e. dielectrics). More specifically, experimental data and simulations following irradiation of fused silica with fs pulses [8] demonstrated an increasing monotonicity of $\Lambda$ with the effective number of pulses at small $N_{eff}$ before a saturation point is reached. The behavior was attributed to the incubation effects that lead to a larger concentration of excited carrier, $N_e$, in the conduction band [6]. As a result, the refractive index of the excited material drops at higher $N_e$ yielding a higher periodicity. Nevertheless, it is recalled that in dielectrics while at moderate excitation levels the periodicity scales as $\sim\lambda_L/n$, at higher excitation levels there is a deviation from this behaviour and the LSFL periodicity moves closer to the laser wavelength [6]. To evaluate whether similar effects occur for polymers, it is noted that, although measurements of the frequencies of the periodic structures have been performed in a different range of effective number of pulses at 10 mJ/cm$^2$ (Fig.4a) and 8 mJ/cm$^2$ (Fig.4b), respectively, it is evident that the enhanced incubation effects at large $N_{eff}$ lead always to a constant periodicity. Thus, a more conclusive investigation is required to determine whether lower excitation levels reached for smaller $N_{eff}$ (Fig.4a,b) result into weak incubation effects which are not sufficient to produce LIPSS of smaller periodicity. A more detailed investigation of the correlation of the excitation level with the size and orientation of LIPSS is performed in the next section via the use of a physical model that describes the role of electrodynamical effects and carrier dynamics.

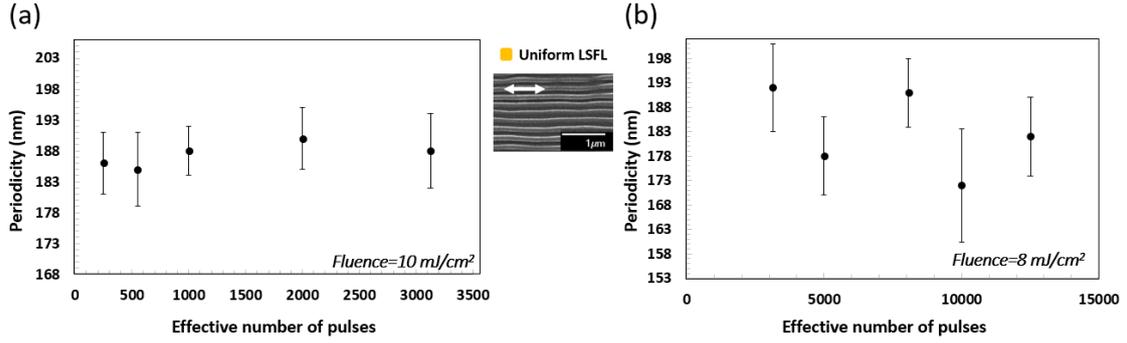

**Figure 4**. **LIPSS periodicity dependence on $N_{eff}$.** Evolution of the LIPSS periodicity as a function of the number of laser pulses at: (a) $F = 10$ mJ/cm² and (b) $F = 8$ mJ/cm². The SEM image shows a characteristic topography that was analysed. The *white* double-headed arrow indicates polarisation direction.

In regard to the dependence of the LIPSS periodicities on the fluence at constant $N_{eff}$, results show (Fig.5a) that $\Lambda$ exhibits a weak increase as $F$ increases for large $N_{eff}$, which also supports the argument that an increased fluence results into a higher excitation level and therefore to a longer periodicity similarly to the behaviour exhibited in dielectrics [6, 8]. To demonstrate that the trend of the monotonicity is maintained regardless the LIPSS uniformity, the analysis illustrated in Fig.5 includes a range of fluences that lead to both non-homogenous and homogeneous LIPSS. More specifically, Fig.5a shows the period evolution over fluence in the case of the homogeneous LIPSS formed on polycarbonate. Here, fluence value ranges between 8 mJ/cm² and 13 mJ/cm². At this fluence range, we observe a, weakly, increasing LIPSS periodicity, namely, from $\Lambda=193$ nm to $\Lambda=220$ nm. By contrast, at higher fluences (and smaller effective number of pulses) and due to the, relatively, large error bars, a conclusive dependence of the LIPSS periodicity on fluence is more difficult despite a rather unchanging trend for $\Lambda$ (Fig.5b). For fluences between 100 mJ/cm² and 400 mJ/cm², non-uniform periodic structures are formed.

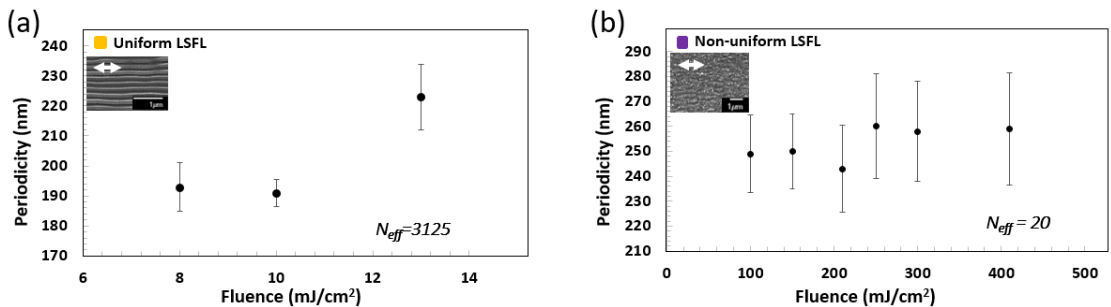

**Figure 5. LIPSS dependence on fluence.** The evolution of the LIPSS periodicity on PC as a function of the fluence at (a) $N_{eff}=3125$ and (b) $N_{eff}=20$.

## *2.1.2 Interpretation of the results*

The observation that the LIPSS period increases and finally saturates has been also reported in previous works performed in other polymeric materials [19]. Interestingly, Rebollar *et* al. [1] state that in polymers, the LIPSS period increases at growing fluence until it reaches a plateau for a value of period similar to the irradiating laser wavelength (scaling as $\lambda_L/n$). Nevertheless, based on the proximity of the periods for the LSFL-I and LSFL-II observed (Fig.2) (not very far from $\lambda_L$), and due to the different origin of the LSFL formation which relates to the excitation level, an interpretation of the mechanisms that lead to either type should be performed by analysing the laser energy deposition and subsequent carrier excitation process. In the following, a theoretical model is employed to describe the formation of LSFL-I and LSFL-II at various excitation levels (i.e. related to the density of the carriers in the conduction band and the laser peak fluence). That model has been used in the past at weak or strong absorbing materials [34].

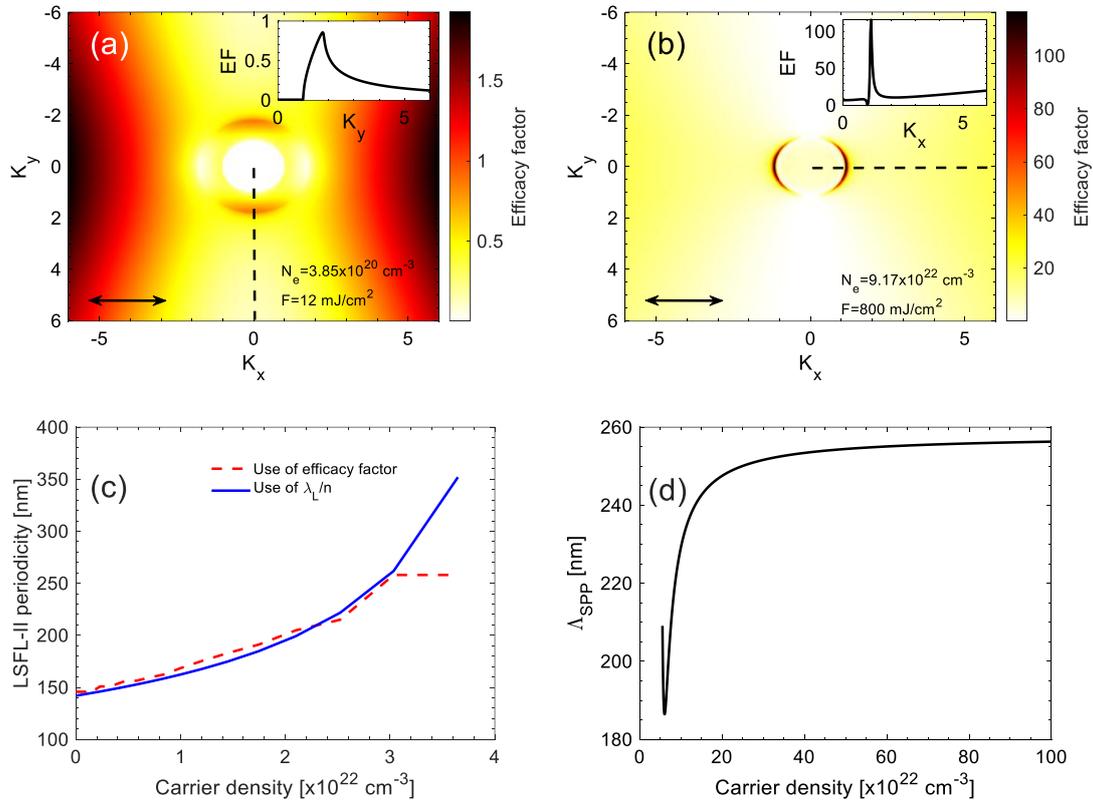

**Figure 6. Modelling LIPSS periodicity for linearly polarised beams.** Efficacy factor map for (a) $N_e=3.85\times10^{20}$ cm$^{-3}$ ($F=12$ mJ/cm$^2$) and (b) $N_e=9.17\times10^{22}$ cm$^{-3}$ ($F=800$ mJ/cm$^2$), (c) LSFL-II periodicity calculated through the employment of the efficacy factor theory and the expression $\lambda_L/n$, (d) Surface plasmon periodicity as a function of the excited carrier density. The *black* double-headed arrow indicates polarisation direction. The dielectric parameter for Polycarbonate (at room temperature) is equal to 1.817+*j*0.02 at $\lambda_L$=258 nm. standard values for the 'shape', *s*, and the 'filling', *f*, parameters (i.e. 0.4 and 0.1, respectively) are used to calculate the efficacy factors and $\Lambda_{SPP}$ (=$\Lambda$).

To interpret the distinctly different structure formation, we correlate the density of the excited carriers $N_e$ with the features of the induced LSFL structures through the use of Sipe's theory [34]. It is recalled that Sipe's theory introduces the interference of the incident laser beam with some form of surface-scattered electromagnetic waves that is excited due to the presence of roughness on the surface. The inhomogeneous energy deposition into the material is computed through the calculation of the product $\eta(\vec{K},\vec{\kappa_i})\times|b(\vec{K})|$; $\eta$ stands for the efficacy with which surface roughness at the wave vector $\vec{K}$ (i.e., normalized wave vector $|\vec{K}|= \lambda_L/\Lambda$, where $\Lambda$ stands for the predicted structural periodicity), induces inhomogeneous radiation absorption, $\vec{\kappa_i}$ is the component of the wave vector of the incident laser beam on the material's surface plane and $b$ represents a measure of the amplitude of the surface roughness at $\vec{K}$. In this work, standard values for the 'shape', $s$, and the 'filling', $f$, parameters (i.e. 0.4 and 0.1, respectively) are used to calculate the efficacy factor and $\Lambda$ at different values of $N_e$ (i.e. $N_e=3.85\times10^{20}$ cm$^{-3}$ (Fig.6a) and $N_e=9.17\times10^{22}$ cm$^{-3}$ (Fig.6b)). The two values are appropriately selected to induce the excitation of surface waves with different propagation directions, resulting in the formation of LSFL-I and LSFL-II structures. According to Sipe's theory, sharp points of $\eta$ appear along the $K_y$ (=$\lambda_L/\Lambda_y$) (inset in Fig.6a) for $N_e =3.85\times10^{20}$ cm$^{-3}$ which indicates that periodic structures oriented parallel to the laser polarization are produced with $\Lambda= \Lambda_x$ (LSFL-II) that is ~147 nm. It is noted that this value is relatively close to the experimental results and similar to the efficacy factor theory-based prediction of the LSFL-II from the expression $\lambda_L/n$ (~143 nm). Additionally, simulation results show a remarkable agreement between the two methods at both low and moderate excitation levels, a trend that has also been observed in dielectrics [6]. In particular, simulation results for the LSFL-II periodicity at different (low and moderate) excitation levels through the employment of either the efficacy factor theory or the expression $\lambda_L/n$ are illustrated in Fig.6c show a remarkable agreement while a noticeable discrepancy is observed at higher $N_e$. Nevertheless, in contrast to the pronounced underestimated periodicity calculated via the use of the expression $\lambda_L/n$ in dielectrics compared to the value derived from the efficacy factor theory [6], results for PC demonstrate an opposite behaviour; furthermore, for $N_e>3\times10^{22}$ cm$^{-3}$ the expression $\lambda_L/n$ leads to non-realistic values for the periodicity of the LSFL-II structures (higher than the laser wavelength) which is attributed to the distinct drop of the refractive index of PC compared to dielectric materials at higher fluence (see Supplementary Materials). Therefore, in the scenario where more uniform LSFL-II are formed, the decrease in refractive index with increasing fluence (caused by the generation of higher $N_e$, see Supplementary Material) is not enough to account for the periodicity of the structures. As a result, the periodicity does not scale as $\lambda_L/n$. Therefore, the efficacy factor theory offers a more accurate description of the increase caused by the scattered electromagnetic surface wave from a defect on the surface of the irradiated solid.

It is emphasised that the dielectric parameter $\varepsilon$ which is used in the calculation of the efficacy factor and the refractive index is $n$ derived from the following expression [29, 35]

$$\varepsilon = 1 + (\varepsilon_{un} - 1)\left(1 - \frac{N_e}{N_V}\right)\frac{e^2 N_e}{m_r m_e \varepsilon_0 \omega_L^2}\frac{1}{\left(1 + i\frac{1}{\omega_L \tau_c}\right)} \qquad (1)$$

In Eq.1, $\varepsilon_{un}$ corresponds to the dielectric parameter of the unexcited material which is derived from the expression $\varepsilon_{un} = (1.817 + j0.02)^2$ [24, 26], $N_e$ and $N_V$ are the carrier densities in the conduction and valence bands ($N_V \sim 10^{23} - 10^{24}$ cm$^{-3}$ for PC), respectively, $m_e$ is the electron mass, $e$ is the electron charge, $m_r = 0.8$, $\varepsilon_0$ is the vacuum permittivity, $\omega_L$ is the laser frequency and $\tau_c = 1.1$ fs stands for the electron collision time [36].

On the other hand, at high excitation levels ($N_e = 9.17 \times 10^{22}$ cm$^{-3}$), sharp points of $\eta$ appear along the $K_x$ ($=\lambda_L/\Lambda_x$) (inset in Fig.6b) which suggests that periodic structures oriented perpendicular to the laser polarization are produced with $\Lambda = \Lambda_x$ close to $\lambda_L$ (LSFL-I).

As seen in Fig.6d, surface waves that lead to LSFL-I predicted from Sipe's theory are expected to be generated from surface plasmon excitation with SP wavelength being $N_e$-dependent. Results coincide with the use of Eq.1 requiring that the real part of the dielectric parameter $\varepsilon$ needs to be smaller than -1 [37] yielding a minimum value for the carrier density equal to $5.5 \times 10^{22}$ cm$^{-3}$ ($F=543$ mJ/cm$^2$) for SP excitation and LSFL-I formation which appears to approach the threshold for the onset of perpendicularly oriented LSFL-I structures (Fig.3a). A similar approach has been followed to explain the periodicity of surface patterns upon the irradiation of dielectrics with IR [8] or mid-IR pulses [7].

A theoretical model developed to correlate the induced carrier densities with the peak fluence was presented in a recent study [29] aiming to derive the carrier dynamics of polycarbonate (PC) at various fluences. According to this model, the carrier densities $N_e=3.85 \times 10^{20}$ cm$^{-3}$ and $N_e=9.17 \times 10^{22}$ cm$^{-3}$, which resulted in two distinct orientations of surface waves (Fig. 6a,b) and led to LSFL-II and LSFL-I, respectively, correspond to maximum densities derived from fluences of $F=12$ mJ/cm$^2$ and $F=800$ mJ/cm$^2$, both of which are comparable to the fluence values used in this study. Experimental results appear to agree sufficiently with the theoretical predictions; more specifically, for $N_e$ $3.85 \times 10^{20}$ cm$^{-3}$ ($F=12$ mJ/cm$^2$), the periodicity of the parallel oriented LSFL is calculated to be ~147 nm while at higher excitation levels $N_e=9.17 \times 10^{22}$ cm$^{-3}$ ($F=800$ mJ/cm$^2$), the periodicity of the perpendicularly oriented LIPSS is estimated to be ~242 nm. Although the calculated the periodicities may be underestimated, primarily due to incubation effects from the increasing number of pulses, the model is aimed to highlight the expected difference in orientation between LSFL-I and LSFL-II for the two fluences.

Simulation results indicate the LIPSS periods increase as the laser fluence becomes larger due to the increase of the carrier density. By contrast, experimental observations illustrated in Figures 4-5 appear to contradict the monotonicity of the theoretical predictions and periodicity remains fluence-independent. It is recalled, though, that in several materials, even in experimental settings, an initial increase in LIPSS period is observed before the values of the periods saturate (see for example, Ref.[11]). Thus, an increase of the LIPSS periods for a range of values of fluence does not contradict the theoretical data. Nevertheless, various other phenomena such as thermal effects could account for saturation mechanisms that prevent further increase of the LIPSS periods. In that case, a revised model is required to take into account relevant processes.

It is important to emphasize that while the efficacy factor theory allows a relatively accurate prediction of LSFL periodicity evolution at both low and higher excitation levels, the methodology is still approximate. This is because it does not take account changes in the surface profile (i.e. deeper profile) as the number of pulses increases. For a more precise estimation, more advanced electromagnetic models should be employed [38-40], however, this is beyond the scope of the current study.

### 2.1.3 Generation of topographies using circularly polarized UV fs pulses

In the previous section, a thorough investigation of the conditions leading to the formation of parallel LIPSS (1D-LIPSS) on PC with linearly polarised UV fs beams was presented. To account for the formation of 2D-LIPSS (or 2D nanostructures), we investigate the morphological features generated with circularly pulses. For this purpose, a parametric study has been performed upon the variation of $F$ and $N_{eff}$. SEM images of surfaces obtained at $F$=250 mJ/cm² are illustrated in Fig.8a showing that at low $N_{eff}$ ($N_{eff}$=17), a topography of high degree of symmetry is produced that comprises 2D nano-protrusions exhibiting an almost perfect circular cross section. As $N_{eff}$ increases (Fig.7a), the uniformity progressively deteriorates and at a relatively large number of effective number of pulses $N_{eff}$ ($N_{eff}$ =200) non-uniform 2D nanostructures are formed. The nanospheres formation can be attributed to the fact the amplitude of the electric vector remains constant for circularly polarized beams and rotates around the propagation direction, causing the irradiated material to heat up, rearrange, and align with the changing polarization directions, resulting in a uniformly distributed pattern [41]. Similar patterns are also produced at lower fluences ($F$=130 mJ/cm$^2$) (Fig.7b).

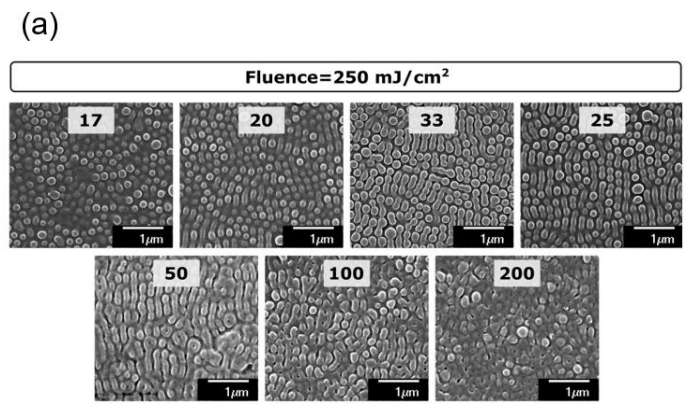

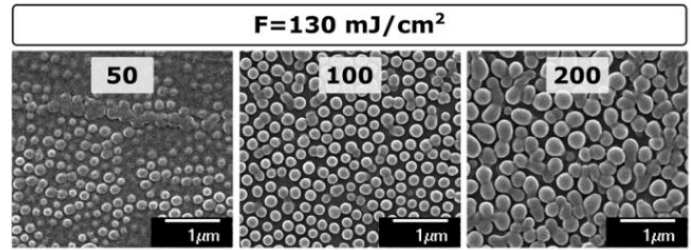

**Figure 7**. **PC structuring with UV circularly polarised beams.** SEM images of PC surface irradiated for various fluences and number of pulses utilising UV circularly polarized fs laser pulses: (a) $F$=250 mJ/cm² (b) $F$=130 mJ/cm². The numbers shown in the images correspond to the effective number of pulses.

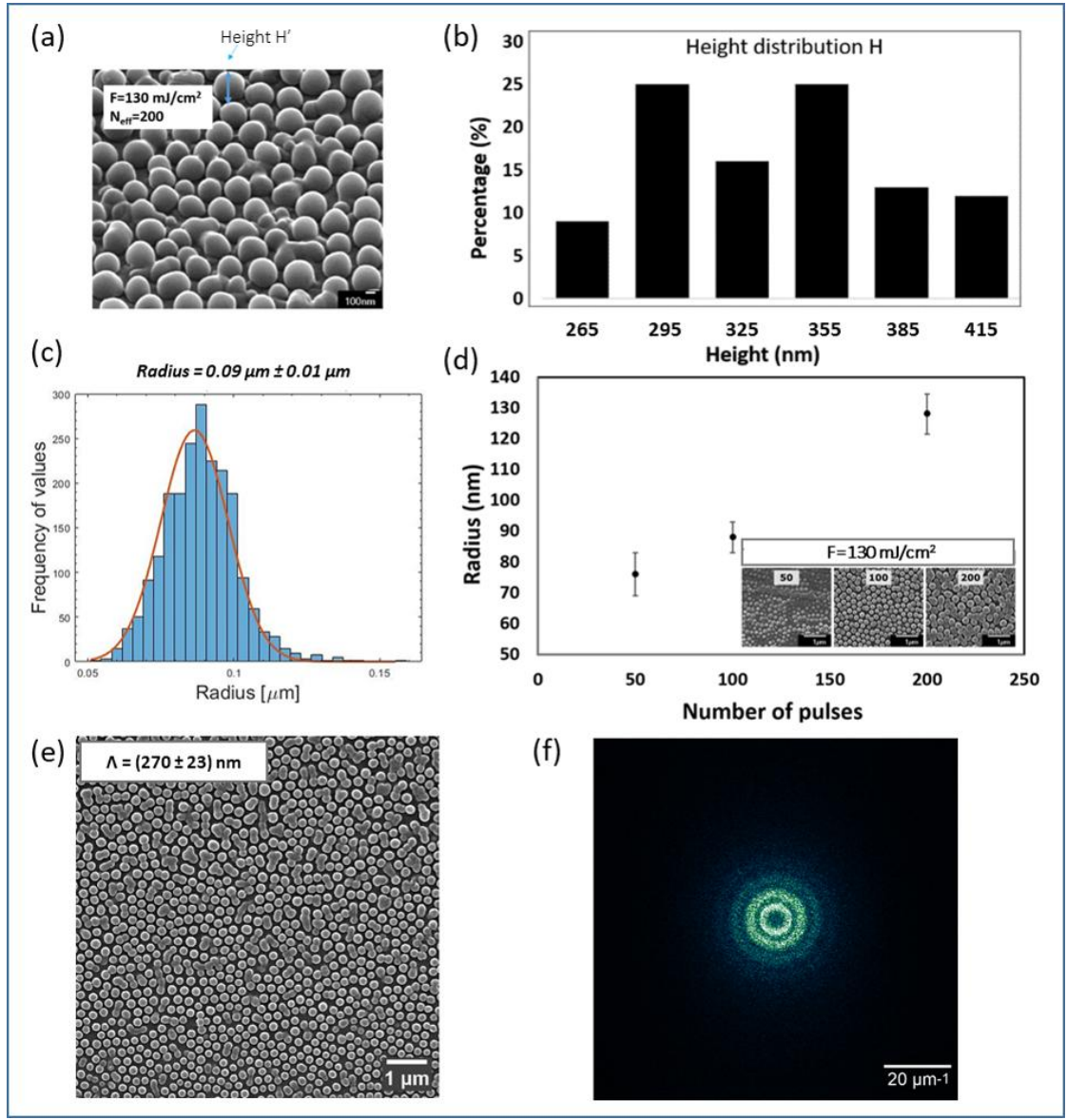

**Figure 8. Features of LIPSS formed with UV circularly polarised beams.** Morphological analysis on the size: period, height and curvature radius of the nanospheres formed on PC. (a) SEM image of nanospheres (tilted by 45⁰), (b) Height distribution of nanospheres analyzed in (a); *H'* corresponds to the height assuming the tilted profile, (c) Curvature radius distribution of SEM image for $F$=130 mJ/cm$^2$ and $N_{eff}$ =100, (d) evolution of radius of curvature size as a function of $N_{eff}$ (values in the inset correspond to the effective number of pulses), (e) SEM image of nanospheres with $F$=130 mJ/cm$^2$ and $N_{eff}$ =100 and (f) the corresponding 2D-FFT analysis.

The analysis of the height distribution (SEM image shown in Fig.8a) leads to the estimation of the mean height of induced nanospheres (~330 ± 34 nm) shown in Fig.8b. On the other hand, the size distribution of the radii of the protruded nanostructures have been evaluated through the employment of home-made Matlab-based Image Analysis techniques (MATLAB version: 9.13.0 (R2022b), Natick, Massachusetts: The MathWorks Inc., 2022) performed on the respective SEM images (see 'Numerical Simulation'). More specifically, for irradiation conditions ($F$=130 mJ/cm$^2$ and $N_{eff}$ =100), the mean value of the radius of the nanospheres is estimated to be 87 ± 12 nm (Fig.8c). A detailed analysis of the data for various number of pulses was conducted indicating that an increasing number of pulses yields protruded structures of larger radius (Fig. 8d). Unlike the observations for linearly polarized beams, the increase of the nanosphere radius when irradiated with circularly polarized beams at increasing the number of pulses can be directly associated to the nature of the polarization. Specifically, circular polarization does not favor a particular direction of laser-material interaction, leading to uniform heating of the material, along with thermal accumulation with each successive pulse. Thus, the increased size of the nanospheres at higher $N_{eff}$ can be due to the occurrence of more intense thermal effects at higher excitation levels which result in a volume expansion of the produced structures. Upon increasing $F$ and $N_{eff}$, the protrusions' mean periodicity increases as well. The average periodicity for $F$=130 mJ/cm$^2$ and $N_{eff}$ =100 is estimated to be equal to $\Lambda$=270 nm (see Fig.8e) which is calculated from a 2D-FFT analysis (Fig.8f). A similar analysis has been conducted in different laser conditions (results not shown).

In previous studies [42-46], the formation of similar 2D-LIPSS patterns using circularly polarized beams in dielectrics or semiconducting materials was reported, where higher peak fluences and a larger number of pulses were employed. In contrast, our findings demonstrate that 2D-LIPSS formation in polycarbonate (PC) can be achieved with relatively low fluences and fewer pulses. This behavior can be attributed to the high absorptivity of polycarbonate at 258 nm which facilitate the occurrence of morphological effects in these conditions. Furthermore, our results showed the ability to control the uniformity of the nanospheres produced on polymers by tuning appropriately the laser parameters (Fig.7); these observations are aimed to contribute to a systematic exploration of the impact of the laser fluence and energy dose on the features of 2D structures following irradiation with UV fs pulses, an area that is yet unexplored [1].

## 2.2 Optical properties of the UV fs laser fabricated structures on PC

A series of reflectance $R$ (Fig.9) and transmittance spectra $T$ (Fig.10) in the ultraviolet (UV)-visible (VIS)-near and infrared (NIR) wavelength ranges were recorded to evaluate the optical response of laser processed PC surfaces. Both spectra are normalised to the maximum detected signal.

Results are illustrated for both the pristine unprocessed and processed PC samples. The measured spectra of pristine PC is comparable to the values for PC reported in previous works [47]. The sequence of peaks occurring at wavelengths above 1100 nm are due to the resonant absorption bands of the carbon-hydrogen bonds [48]. In all cases, the UV-VIS-Near IR optical spectra exhibit remarkable similarity between non-irradiated and irradiated samples (Fig. 10) indicating that PC preserves its optical properties even after processing.

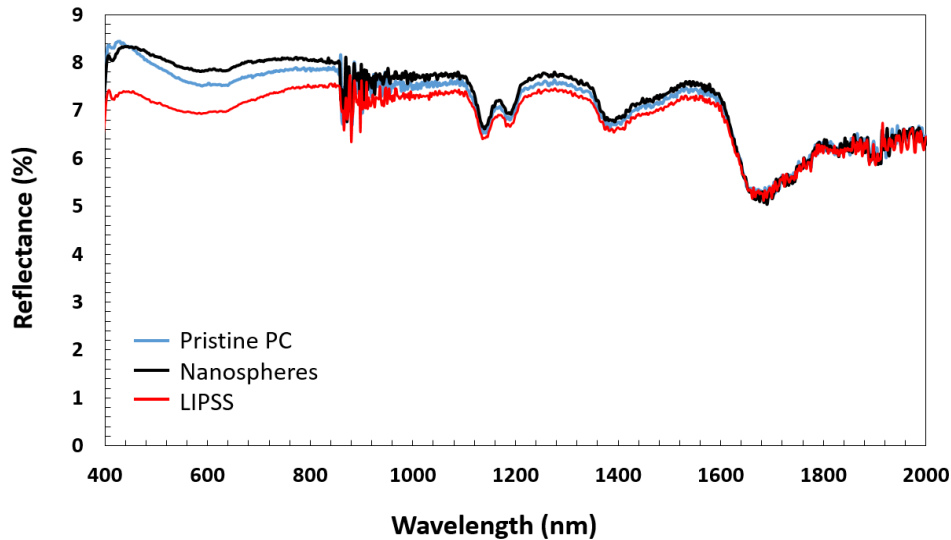

**Figure 9. UV-VIS-Near IR reflectance spectrum.** Spectrum for unprocessed pristine PC target and PC structured with homogeneous LSFL-II or nanospheres.

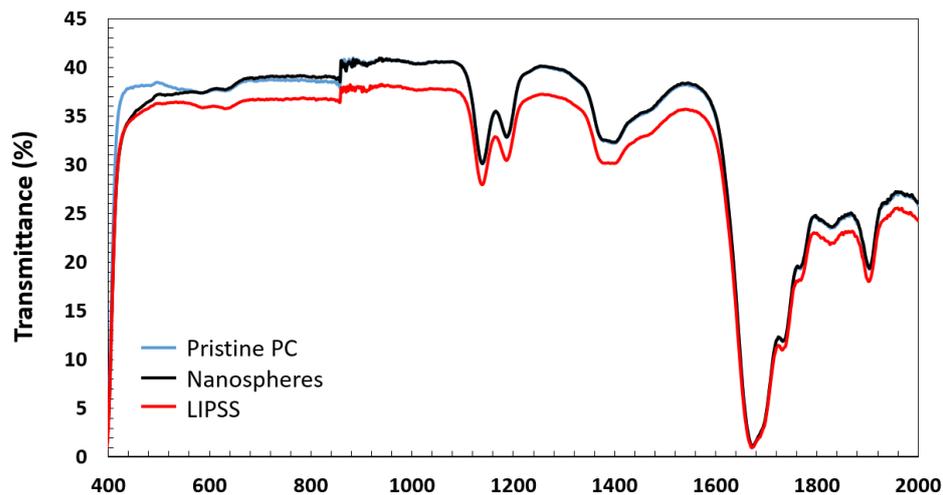

**Figure 10. UV-VIS-Near IR transmittance spectrum.** Spectrum for unprocessed pristine PC target and PC structured with homogeneous LSFL-II or nanospheres.

## 2.3 Wetting properties of the UV fs laser fabricated structures on PC

To investigate the wetting behaviour of the laser structured PC surface with the pristine surface, a series of contact angles (CA) measurements have been conducted. Five sets of measurements were performed for each sample, and the CA was calculated as the average of these measurements. The deviation from the initial value was approximately 2-3% of the CA. The CA of the pristine PC has been measured to be $81.4° \pm 0.4°$. The first CA measurements of the processed samples (images in first row in Fig. 10) were carried out three days after irradiation. It is observed that no remarkable changes in the CA value appear between the pristine and the laser-processed PC surfaces. This indicates that following laser irradiation neither the surface morphology nor the chemical bond changes affect the surface wettability. The test immediately after the laser processing allows for the evaluation of how the laser treatment has modified the surface's properties, such as roughness and chemistry, which directly influence wetting behavior. By contrast, CA measurements conducted for all surfaces after, approximately, one year (images in second row in Fig. 11), revealed that all post-processed surfaces, present a more hydrophilic behaviour. This behaviour is probably attributed to additional chemical changes on the surface since the sample has been exposed to ambient air conditions. The wettability measurements performed one year after laser processing were intended to evaluate the long-term stability and durability of the laser-induced surface modifications. This is particularly relevant for assessing the practical applicability of the technique, as stable wettability is often a critical requirement for real-world applications. Such long-term testing is especially important for ensuring the technique's applicability in environments where the material may be exposed to factors such as UV radiation, moisture, or temperature fluctuations. Nevertheless, more investigation is required to evaluate the influence of a potential chemical modification which is not within the scope of the present work. While the stabilization period for laser-treated materials is often shorter (e.g. within a few days or weeks), the measurements conducted one year later in our study were aimed at confirming that the wettability properties remain consistent and effective over extended time scales, which further supports the technique's suitability for practical applications.

It has been reported that the wettability of laser nano-structured polymer samples depends on both the specific polymer itself and the applied laser irradiation conditions [1]. For example it was reported that for PET film samples irradiated with UV laser pulses, the CA of water decreases, indicating that the surface displays a more hydrophilic character after irradiation [1]. Furthermore, it has been reported that UV fs laser irradiation of polyethersulphone polymers at low fluences improves the surface hydrophilicity [1, 49]. These reports agree with the outcome of this study. By contrast, in the case of irradiation conditions leading to high-aspect ratio polymer structures it is found that the surface hydrophobicity progressively increases with the number of pulses, since the structure depth increases as well [30, 49].

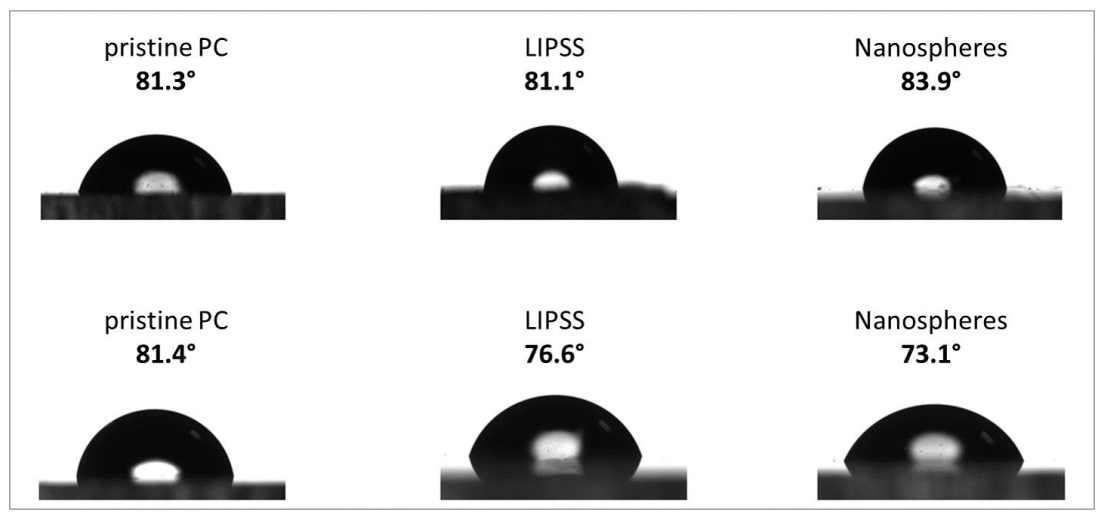

**Figure 11. Wetting properties for structured PC.** Contact angle measurements of pristine PC surface, of irradiated PC surface, structured with homogeneous LSFL-II structures and structured with protruded nano-structures (nanospheres). The measurements (at the second row) were repeated one year after the laser-irradiation.

A more detailed analysis of the correlation of the morphological features of the patterned material with its wetting properties requires the evaluation of the role of the aspect ratios (i.e. length-to-width, length-to-height, width-to-height) to these properties [50]. Such as investigation would necessitate precise measurements of the heights of the ripples via an Atomic Force Microscopy (AFM)-based analysis as it is known that SEM data are not capable to provide information about the amplitude of the laser-induced topography. Despite the evident significance, an association of the particular features of the nanostructured surface with the wetting properties is not within the scope of the present study; thus, only the effect of non-patterned and patterned (with LSFL-II and protrusions) surfaces is, only, discussed in this work.

It is emphasized that the investigation of the comparison of the wetting and optical properties of nonpatterned and nanostructured topographies for polymers irradiated with UV fs pulses is centred on the consideration on patterns comprising of either LSFL-II or protrusions. Although a more conclusive analysis should, also, involve the response of surfaces covered with LSFL-I, limitations related to the conditions required for the fabrication of this type of structures indicate the need for further investigation. More specifically, higher fluence or number of pulses are necessitated to fabricate LSFL-I structures which are expected to lead to an increase of the chemical changes of the irradiated material; this alteration will complicate the identification of the correlation of the topography features with the observed properties. Furthermore, regarding the morphological features (and potentially the optical and wetting properties) of the processed material, our results indicate that the formation of LSFL-I structures requires higher excitation levels than LSFL-II, which may result in deeper patterns. However, as discussed in the main manuscript, a more detailed investigation of thermal effects, which are critical in determining LIPSS depth, is beyond the scope of this study.

## 3. Conclusions

Surface structuring of PC was systematically investigated upon irradiation with linearly and circularly polarised UV fs laser pulses. Our analysis demonstrated that while the former polarisation leads to the formation of 1D structures (LSFL-I and LSFL-II of similar periodicity and orientation as in dielectrics), circularly polarised beams induce 2D patterns comprising nanospheres of comparable sizes with those in dielectric materials, however, at lower fluences and energy doses. In both polarization cases, it was shown that the morphological features, such as the homogeneity, orientation, periodicities and sizes of the structures are significantly affected by the enhanced energy absorption exhibited by PC surface at 258 nm. In particular, our results showed the formation of topographies featuring highly uniform nanospheres, which contrasts with the less uniform structures observed in previous studies involving the irradiation of polymeric materials with circularly polarized beams. To account for the various pattern formation, a correlation of the characteristic structure sizes with the energy dose was analyzed, using a model based on Sipe's theory, and discussed. Theoretical results at low, moderate and high excitation levels predict the orientation and size of the periodic structures of PC that resemble those of dielectric materials despite the enhanced absorptivity and large extinction coefficient of PC compared to transparent solids. A correlation between the excitation level (i.e., carrier densities) and laser energy is presented using a theoretical model for the first time in polymeric materials. The analysis of the optical and wetting properties reveals that the material's response remains unchanged after processing. It can be concluded that the employment of UV fs pulses of different polarisation states can be an efficient tool for the fabrication of highly uniform submicron sized structures on polymeric samples.

## Materials and Methods

### Experimental setup

Commercially available, polycarbonate (PC) plates, purchased from Kintec company, with a thickness of 0.5 mm are used as samples. After irradiation, the sample surfaces were cleaned using ultrasonic cleaning for five minutes by submersion in 70% ethanol and dried with compressed air. The samples were irradiated using the fourth harmonic

(258 nm) of an Yb:KGW pulsed laser source (PHAROS SP) producing femtosecond laser pulses with 170 fs pulse duration and 1 kHz repetition rate. Irradiation of the surface of PC is performed at normal incidence and in ambient air conditions. The polarization of the laser beam is controlled to be linear or circular, while the pulse energy is controlled via a half-wave plate and a linear polarizer. The beam is focused on the material surface by a plano-convex, achromatic spherical lens with a focal length of 100 mm, resulting in a focal spot with a $1/e^2$ beam diameter of $d = 11$ µm, which was measured by a CCD camera. The samples were mounted on a motorized XYZ linearly translation stage and placed normal to the incident laser beam. Line or area scans were produced at variable velocity values. The sample is translated parallel to polarization axis. The moving speed of the translation stage (i.e. the speed of the sample) is correlated with the effective number of pulses irradiating the surface by the following equation (Eq.2), where $\omega_0$ is the focal spot radius (at $1/e^2$), $f$ is the repetition rate of the laser, $v$ is the speed and $y$ is the distance between two sequence lines in the $y$-axis. The fluence of irradiation is determined by measuring the laser energy and by calculating the area of the laser spot on the focal plane. A schematic of the experimental setup is shown in Fig.12

$$N_{eff} = \frac{\pi\omega_0^2 f}{vy} \tag{2}$$

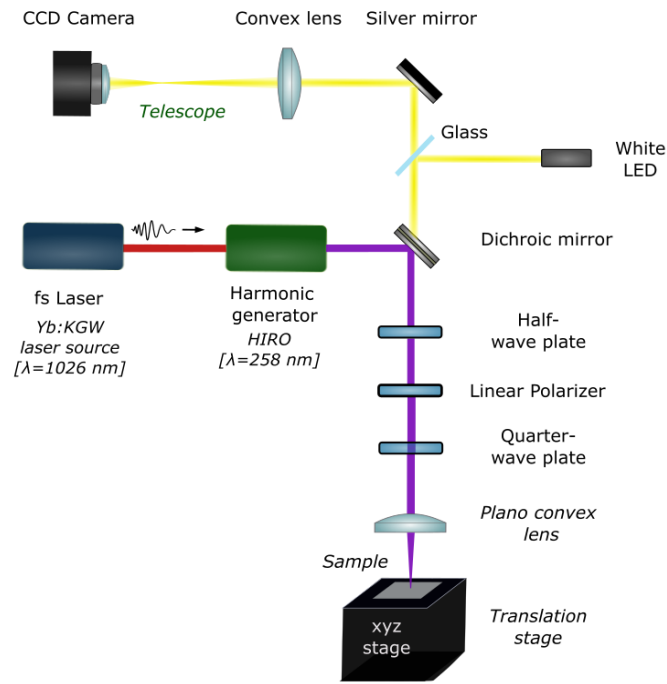

**Figure 12. The experimental setup used for the irradiation experiment**

The morphologies of the laser-fabricated structures were visualized by a field-emission Scanning Electron Microscope, SEM (JEOL JSM-7000F, JEOL Ltd., Tokyo, Japan).

All measurements of the morphological features of surface structures were performed by a 2D-FFT analysis of the corresponding SEM images using Gwyddion (http://gwyddion.net/), a free and open-source software. Optical and wetting properties of structured surfaces are conducted using optical spectroscopy and contact angle (CA) measurements respectively. In particular, the UV/VIS reflection and transmission spectra of the samples, in the range of 400-2000 nm, were recorded using a Perkin Elmer-Lambda 950 UV/VIS spectrometer (PerkinElmer Inc., Waltham, MA, USA). The wetting properties of the fabricated surfaces were evaluated via water contact angle (CA) measurements performed by using the Data Physics OCA 20 system. In particular, distilled water drops of 4 μL were deposited on each surface tested and the mean value of the CA was obtained from five different measurements.

**Numerical simulation**
The efficacy factor-theory based simulations performed to calculate the orientation and the periodicity of the scattered electromagnetic waves that are the precursors of the LSFL structures are conducted via the development of a numerical code developed in Matlab (MATLAB version: 9.13.0 (R2022b), Natick, Massachusetts: The MathWorks Inc., 2022) assuming standard expressions presented in detail in previous reports [51, 52]. The calculation and the dispersion of the curvature radius distribution (Fig.9c) has been calculated via an Image Analysis-based algorithm developed in Matlab. The analysis of each SEM image comprising nanospheres was based on setting an intensity threshold and considering an appropriately selected minimum and maximum value of the radii of the circular objects (in 2D) so that very small (i.e. of low significance) objects are ignored while the periphery of the rest of the objects is precisely defined from the algorithm.

**Acknowledgements**
The authors would like to acknowledge the support provided by A. Manousaki for the SEM characterization. This research was supported by the EU-Horizon 2020 Nanoscience Foundries and Fine Analysis (NEP) Project (Grant agreement ID: 101007417).

**Author Contributions**
M.V. conducted experiments and led data analysis; N.P conducted experiments and contributed to data analysis; S.M. contributed to data analysis; G.T. developed the theoretical model, software and algorithm, and correlated theory with experiments; E.S. conceived the idea behind the experiments. All authors contributed to manuscript writing.

**Conflict of interest**
The authors declare no competing interests.

*Supplementary Material for*

# Fabrication of highly uniform laser-induced periodic structures on polycarbonate via UV femtosecond pulses

Matina Vlahou[1,2], Nektaria Protopapa[1,3], Stella Maragkaki[1], George D. Tsibidis[1] and Emmanuel Stratakis[1,3,*]

[1] Institute of Electronic Structure and Laser (IESL), Foundation for Research and Technology (FORTH), N. Plastira 100, Vassilika Vouton, 70013, Heraklion, Crete, Greece

[2] Department of Materials Science and Technology, University of Crete, 71003 Heraklion, Crete, Greece

[3] Department of Physics, University of Crete, 71003 Heraklion, Crete, Greece

*stratak@iesl.forth.gr

# 1. Surface processing of PC with UV fs pulses - Damage threshold of polycarbonate

Laser-induced structuring of a material's surface is the result of a sequence of complex physical processes involving laser energy absorption by the free electrons generated on the sample surface, energy transfer from the electron system to the lattice and eventually the occurrence of thermal or mechanical modification effects. In particularly, the evaluation of the laser-induced damage threshold of the irradiated material is a subject of great interest in laser-materials processing. Here, we use a conventional experimental approach for defining the damage threshold of the PC material under study, based on its irradiation with a single laser pulse at various fluences and the subsequent measurement of the diameter of the respective ablated spots. The spatial fluence distribution at the focal point of the incident Gaussian beam is given by [53]

$$D^2 = 2\omega_0^2 \ln\frac{F}{F_{thr}} \qquad (1)$$

where $D$ stands for the ablated spot diameter, $\omega_0$ corresponds to the beam radius at $1/e^2$ intensity while $F$ and $F_{thr}$ are the laser and damage threshold (peak) fluences respectively. To calculate the damage threshold, the average of $D^2$ is plotted as a function of the laser fluence $F$ and is shown in Fig.SM1. Upon fitting these data with Eq.1, the calculated ablation threshold for $D^2=0$ is estimated to be, approximately, equal to 0.28 J/cm².

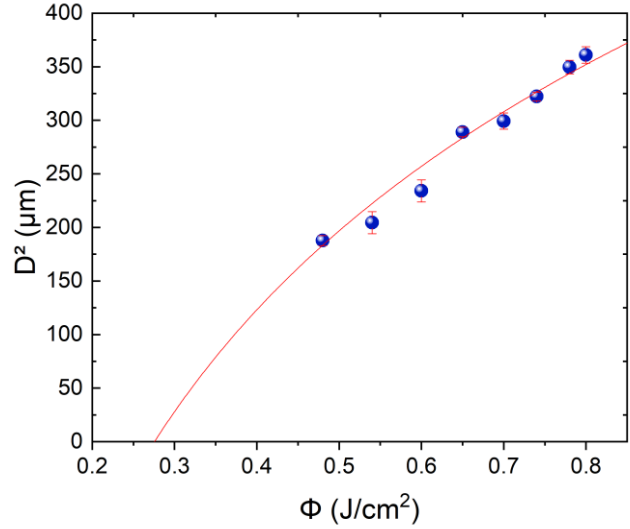

**Figure SM1**. **Procedure to evaluate the Damage Threshold.** The square of the ablated surface diameter as a function of the laser fluence.

It is well known that the damage threshold is dependent on the number of pulses [51, 54, 55]. Incubation effects, caused by an increasing number of pulses, affect the variation in the damage threshold, resulting in a sharp decrease to a value much lower than the single-shot damage threshold (~40%-70% smaller as reported in previous works). A saturation in the reduction of the damage threshold is typically reached after a small number of pulses. In particular, in Ref. [51], it is mentioned that a reduction in the damage threshold due to incubation effects occurs, also, in polymeric materials.

In this study, we use combinations of fluence ($F$) and number of pulses ($NP$) values specifically, ($F$ = 500 mJ/cm² at $NP$ = 1) and ($F$ = 8 mJ/cm² at a minimum $NP$ of 3125, a three-order magnitude difference in $NP$) to demonstrate the adequacy of the fluence in causing material damage.

On the hand, it has been reported (see Ref.[1]) that LIPSS formation in polymers occur at fluence values well below the ablation threshold of the material.

In Figure SM1a,b,c, SEM images of ablated-damaged areas of irradiated spots on PC at $F$= 0.8 J/cm², $F$= 0.48 J/cm², and $F$=0.42 J/cm² are illustrated. An analysis of the SEM images shows that two distinct affected areas are observed on the produced spots, characterized by an inner and outer ring. The radii of the two rings are indicated by a *green* and *blue* arrow, respectively). At higher fluences ($F$= 0.8 J/cm² and $F$= 0.48 J/cm²) the inner ring boundary is more distinct while the outer ring boundary is blurred and nearly absent; therefore inner ring's diameters have been measured and used in all cases to determine the ablated spot diameter, D except from spots produced at fluences lower than $F$= 0.48 J/cm² (Figure 1c).

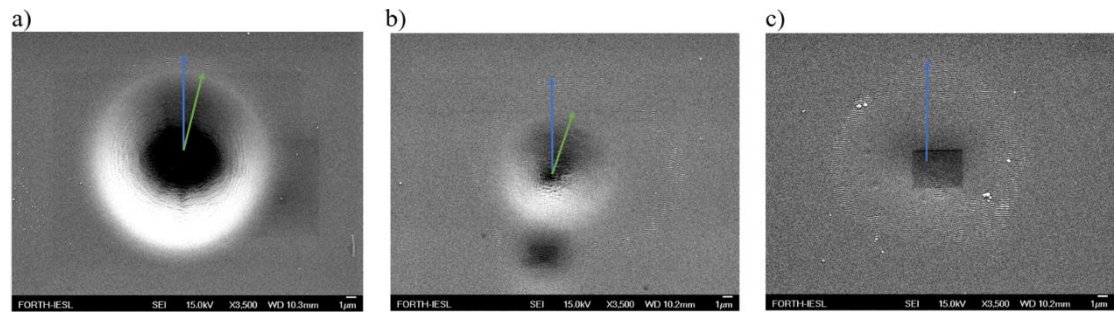

**Figure SM2. SEM images of damaged laser spots on PC**: for a) $F$= 0.8 J/cm², b) $F$= 0.48 J/cm² and c) $F$=0.42 J/cm².

## 2. Refractive Index at different excitation levels and collision time

The refractive index (*n*) dependence on the excitation level (i.e. excited carrier density) is shown in Fig.SM3. It is shown that the electron collision time influences the dependence of the refractive index. Results illustrate the variation of the refractive index at two different values of the electron collision time ($\tau_c$=1.1 fs and $\tau_c$ =0.4 fs). The value used in the calculations in this work is $\tau_c$=1.1 fs as there is an agreement with experimental data (results for $\tau_c$ =0.4 fs were illustrated for fused silica in a previous report [6]). The calculations show that at low excitation levels (i.e. small carrier density), the refractive index remains almost constant while it drops abruptly at higher excitation. Once it reaches a minimum value, *n* begins to rise again, and this increase is typical when the material exhibits metallic characteristics. It is noted that the *dashed blue* (Fig.SM3b) distinguishes the regions where the material stops to behave as a polymer and it attains a metal character (for $N_e^{(m)}$=4.27×10$^{22}$ cm$^{-3}$ which occurs when the real part of the dielectric parameter is equal to zero [36]).

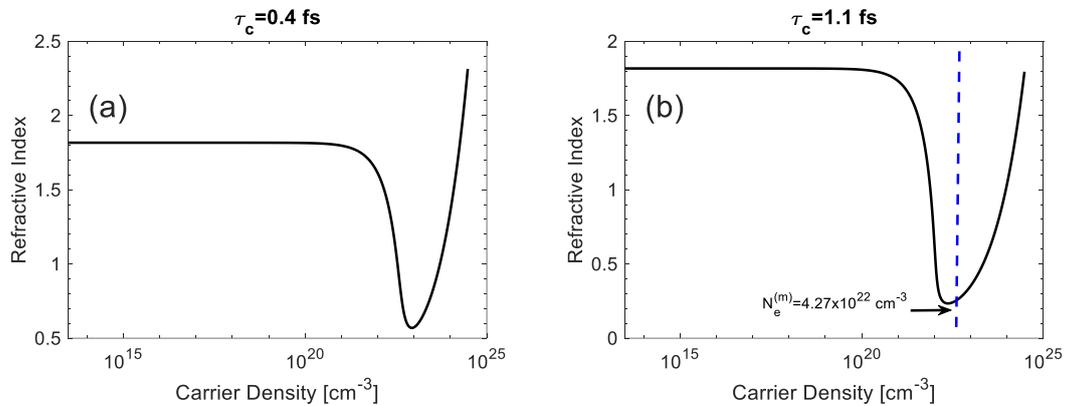

**Figure SM3**. **Refractive index vs Carrier density.** Results are shown at two different values of the electron collision times ($\tau_c$=0.4 fs (a) and $\tau_c$ =1.1 fs (b)). The *blue* dashed lines in (b) indicate the value at which the material starts to behave as a metal ($N_e^{(m)}$=4.27×10$^{22}$ cm$^{-3}$).

The extinction coefficient as a function of the carrier density is illustrated in Fig.SM4. Results show the large increase of the extinction coefficient at large carrier densities (especially, when the material attains a metal behaviour.

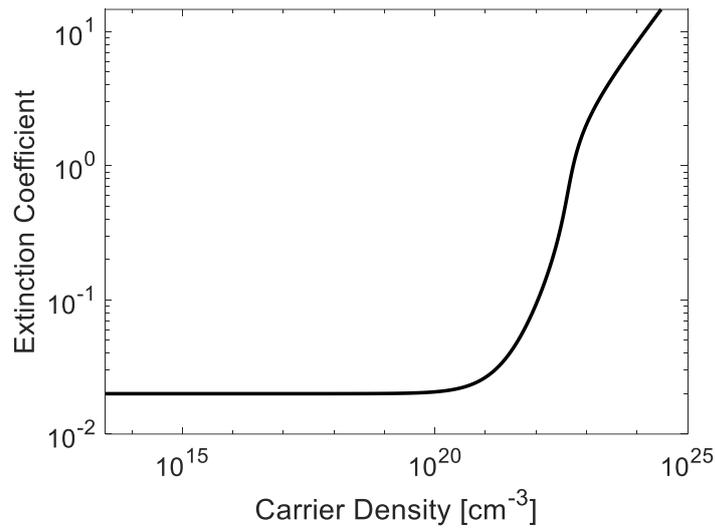

**Figure SM4**. **Extinction coefficient vs Carrier density.** Results are shown for $\tau_c$ =1.1 fs.

## 3. Relation of fluence with maximum carrier density

In the main manuscript, the periodicity of the LIPSS (LSFL-I and LSFL-II) was illustrated as a function of the maximum density of the excited carriers. To relate the excitation level with the laser fluence, a theoretical model [29] was used to derive the dependence of the maximum carrier density with the laser fluence (Fig.SM5).

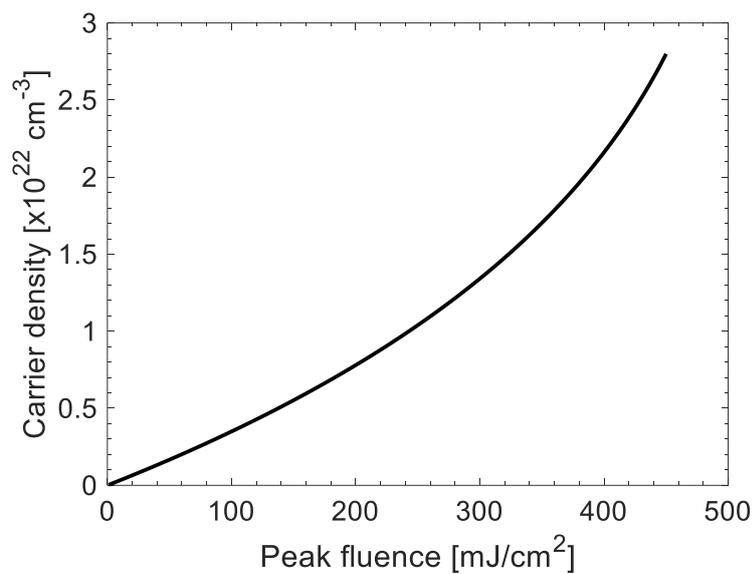

**Figure SM5**. Dependence of carrier density with laser fluence following irradiation of PC

with the laser source.

## 4. Uniformity of the periodic structures

An analysis has been performed to clarify the criteria under which LIPSS and nanospheres are characterized as either uniform or non-uniform. More specifically, we have performed a 2D-FFT analysis of the spatial frequency distribution on patterned areas (see region in Figures SM6,7 outlined by the purple lines). Below, we provide a discussion of the distinction between 'uniformity' and 'non-uniformity' of the periodic topographies (also included in the Supplementary Material)

1. LIPSS are illustrated in Figure SM6a-d using a linearly polarized beam with their corresponding 2D-FFT maps. In the case of a 'uniform' LIPSS covered topography (Figure SM6b), the pattern exhibits a regular, periodic arrangement projected on the 2D-FFT map (Figure SM6d) by sharp and well-defined peaks at specific spatial frequencies that correspond to the structure's periodicity. By contrast, for 'non-uniform' patterns the periodicity of the structures varies across the surface (Figure SM6a) and the 2D-FFT (Figure SM6c) shows a more diffused or irregular frequency spectrum. Instead of sharp peaks, the 2D-FFT exhibits broader features and a more complex arrangement of peaks, indicating the varying spatial frequencies present in the pattern.

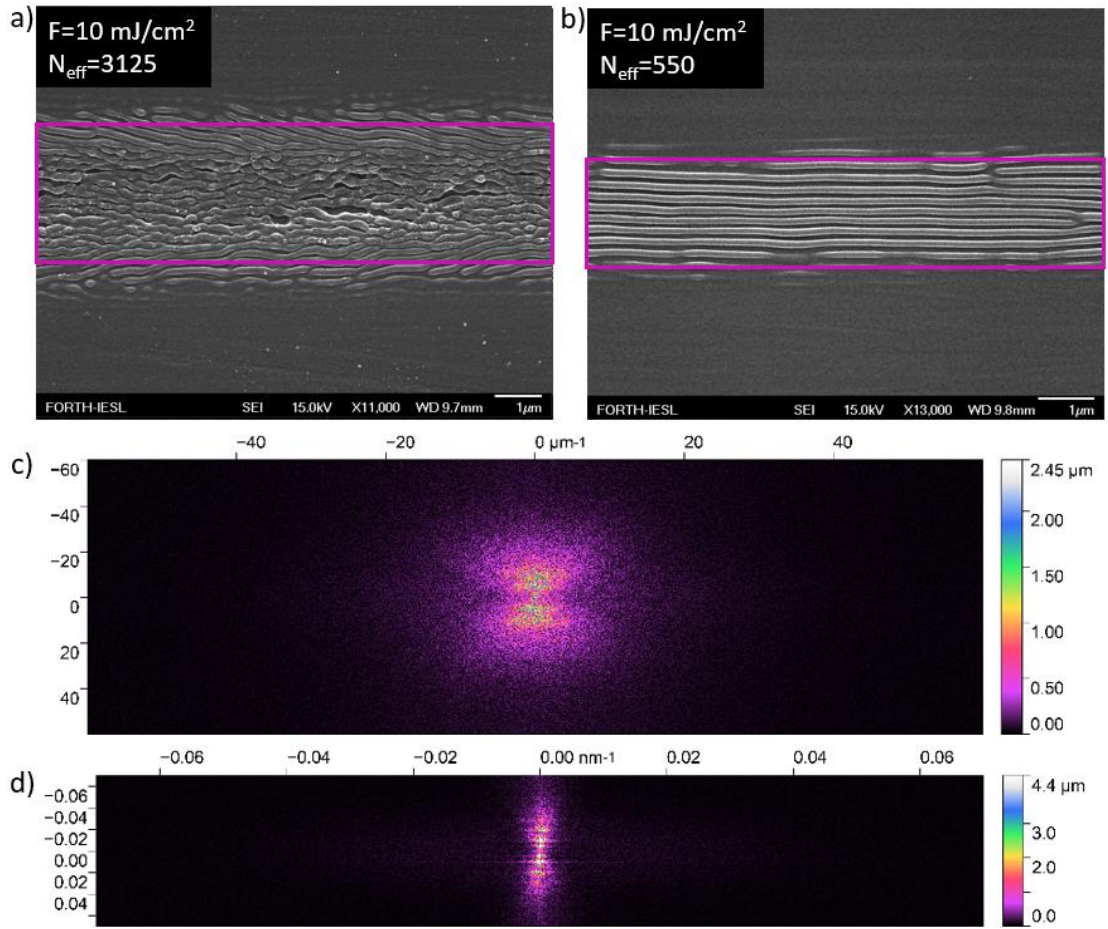

**Figure SM6**. SEM images of produced structures and corresponding 2D FFT analysis for the case of linear laser polarization used. a) SEM of non-uniform LIPSS, b) SEM of uniform LIPSS, c) 2D FFT analysis of non-uniform LIPSS d) 2D FFT analysis of uniform LIPSS

2. Nanospheres are illustrated in Figure SM7a-d using a circularly polarized beam with their corresponding 2D-FFT maps. In contrast to the case of irradiation with linearly polarised beams, when a circularly polarized beam is used to induce Laser-Induced Periodic Surface Structures (LIPSS), the resulting LIPSS pattern tends to be more isotropic. This means that the structures (nanospheres) are distributed in all directions, with no specific orientation (Figure SM7b). The 2D-FFT shows a radial symmetry, meaning that the frequency components are spread uniformly around the center of the 2D-FFT plot (Figure SM7d). This is because circular polarization creates structures that are equally spaced in all directions. The FFT map exhibit circular rather than sharp peaks along a specific axis. This reflects the isotropic nature of the pattern and indicates that the periodicity of the LIPSS is the same in all directions. By contrast, the production of non-uniform structures (Figure SM7a) reflects on the formation of non-concentric/radially arranged peaks on the 2D-FFT as shown in (Figure SM7c).

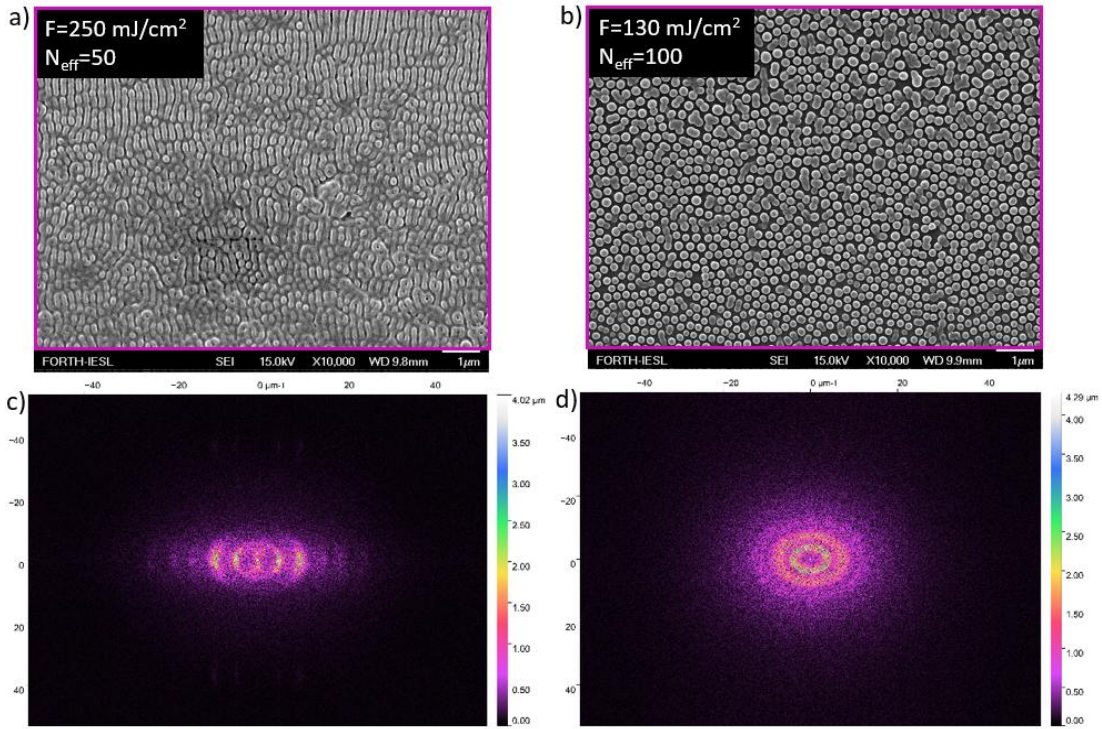

**Figure SM7.** SEM images of produced structures and corresponding 2D FFT analysis for the case of circular laser polarization used. a) SEM of non-uniform nanospheres, b) SEM of uniform nanospheres, c) 2D FFT analysis of non-uniform nanospheres and d) 2D FFT analysis of uniform nanospheres